\documentclass[preprint,
 amsmath,amssymb,
 aps,
]{revtex4-1}
\usepackage{graphicx}
\usepackage{natbib}
\usepackage{dcolumn}
\usepackage{bm}
\usepackage[squaren,Gray]{SIunits}
\begin{document}

\title{Remote plasmon--induced heat transfer probed by the electronic transport of a gold nanowire}

\author{M.-M. Mennemanteuil}
 \affiliation{Laboratoire Interdisciplinaire Carnot de bourgogne, CNRS-UMR 6303, Universit\'e Bourgogne Franche-Comt\'e, 21078 Dijon, France}
 
\author{M. Buret}
\affiliation{Laboratoire Interdisciplinaire Carnot de bourgogne, CNRS-UMR 6303, Universit\'e Bourgogne Franche-Comt\'e, 21078 Dijon, France}

\author{N. Cazier}
\affiliation{Laboratoire Interdisciplinaire Carnot de bourgogne, CNRS-UMR 6303, Universit\'e Bourgogne Franche-Comt\'e, 21078 Dijon, France}

\author{M. Besbes}
\affiliation{Laboratoire Charles Fabry, Institut d'Optique Graduate School, CNRS, Universit\'e Paris-Saclay, 91127 Palaiseau, France.}

\author{P. Ben-Abdallah}
\affiliation{Laboratoire Charles Fabry, Institut d'Optique Graduate School, CNRS, Universit\'e Paris-Saclay, 91127 Palaiseau, France.}

\author{G. Colas-Des-Francs}
\affiliation{Laboratoire Interdisciplinaire Carnot de bourgogne, CNRS-UMR 6303, Universit\'e Bourgogne Franche-Comt\'e, 21078 Dijon, France}

\author{A. Bouhelier}
\affiliation{Laboratoire Interdisciplinaire Carnot de bourgogne, CNRS-UMR 6303, Universit\'e Bourgogne Franche-Comt\'e, 21078 Dijon, France}

\date{\today}

\begin{abstract}
We show in this paper that the heat generated by the optical excitation of resonant plasmonic antennas and diffusing along a simple glass/air interface disturbs the electron transport of a nearby conductive element. By probing the temperature-dependent resistance of a gold nanowire $R_{\rm nw}(T)$, we quantitatively analyze the impact of a resonant absorption of the laser by the antennas. We find that the temperature rise at the nanowire induced by the laser absorption of a distant nanoparticle may exceed that of a direct illumination of the nanowire itself. We also find that a global temperature calibration underestimates the heat generated locally by the laser.  The temperature deduced from resistance variations are verified by numerical simulations with a very satisfactory agreement.
\end{abstract}

\pacs{65.80;-g, 72.25.Jf, 73.20.mf}
\maketitle


\section{\label{sec:level1}Introduction}
The optical interaction cross-sections of gold nanostructures play a central role for improving the emission characteristics of light-emitting devices~\cite{Cho2011} and are at the core of highly-sensitive nanoscale biosensors~\cite{Nusz2008,Nanostars2010,Nyagilo2015}. This plasmonic technology is gaining ground in the development of novel optoelectronic components~\cite{Ward2010,Stolz2014,hotelectron} and is now utilized to engineer thermal landscapes~\cite{Baffou2010,Sanchot2012,Baffou2013,Baffou2014}. For instance,  heat resulting from the resonant absorption of an incident radiation at the plasmon frequency is exploited in photothermal cancer therapy~\cite{Halas2001,Dickerson2008} and is applied to initiate local catalytic reactions~\cite{Baffou2010,Baffou2014,halas2014}. A quantitative measurement of the local elevation of the temperature is typically obtained by measuring the subtle change of the refractive index of the surrounding medium~\cite{Baffou2012}, the anisotropy of polarization of a fluorescent specie~\cite{Baffou2009} or its temperature-dependent lifetime~\cite{Coppens2013a}. These measurements rely on a medium embedding the nanoheaters and an estimation of the temperature is restricted to a few hundred nanometers corresponding to the water or polymer shell surrounding the nanoparticle~\cite{Kotaidis2005,Govorov2006,Volkov2007}. The temperature elevations inferred from these investigations suggest that the thermal relaxation to the substrate is a factor that should not be neglected~\cite{BenAbdallah11, Carlson2012,Desiatov2013}, in particular when operating plasmonic optoelectrical components. While bolometric and thermocouple measurements are used for temperature at the nanoscale~\cite{Weeber2011,Herzog2014}, the heat generated by optical pumping of an electrically connected plasmonic device may disturb the flow of current by increasing the device's resistance~\cite{Herzog2014} and is a source of undesired thermo-voltages~\cite{Stolz2014,Halas2015,Leonard2015}. 

In this paper, we investigate the impact of a remote plasmon-induced heating on the electronic transport of a Au nanowire. By probing the change of resistance of the nanowire, we quantify the local temperature variation resulting from the optical absorption of distant Au nanodisks. We find that for a resonant photothermal excitation, the electron transport in the nanowire senses the heat produced by nanodisks displaced as far as 6~$\mu$m~away. We also find that the temperature rise at the nanowire induced by the optical pumping of a remote nanoheater may exceed that of a direct illumination of the nanowire itself.

\section{\label{sec:level1}Experimental methods}
\subsection{\label{sec:level1}Nanofabrication}

\begin{figure}[h]
\includegraphics[width=15cm]{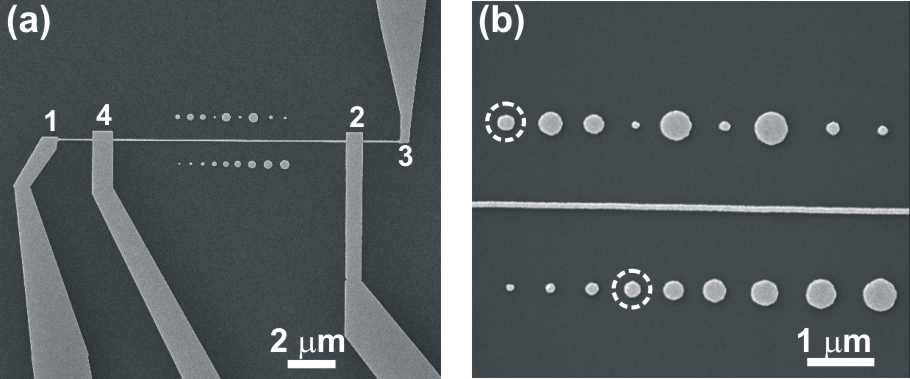} 
\caption{(a) Scanning electron image of a contacted Au nanowire decorated by adjacent Au nanodisks in a 4-probe resistance measurement. (b) Close-up image of the gold nanodisks placed nearby the nanowire. At the upper row, the diameters of the nanodisk are randomly arranged. At the bottom row, the diameters are increasing from left to right. The circled nanodisks are resonant with the laser wavelength.  }
\label{SEM}
\end{figure}

The gold nanowires and nanodisks are fabricated by electron beam lithography on a standard glass coverslip. The nanowire has a nominal length of 10~$\mu$\meter~and a width of 75~n\meter. The nanodisks are placed adjacent to the nanowire with varying diameters ranging from 90~n\meter~to 410~n\meter. The structures are constituted of a 5~nm thick Cr adhesion layer and a 60~nm thick layer of Au, which are both deposited by thermal evaporation. The nanowire is electrically connected to a set of macroscopic Au electrodes produced by standard photo-lithography and thermal deposition. Figure~\ref{SEM}(a) shows a scanning electron micrograph of a typical sample used in this work. It consists on four macroscopic electrodes connecting the nanowire and used to measure its resistance in a 4-probe configuration. The nanowire and the nearby nanodisks are readily seen at the center of the image. Figure~\ref{SEM}(b) is a close-up image of the nanodisks. At the bottom of the nanowire, we design nanodisks with increasing diameters from left to right whereas at the top of the nanowire, the same set of diameters is randomly ordered. The center-to-center distance between the nanowire and the nanoparticle is constant at 1~$\mu$\meter~and the interparticle edge separation is 500~n\meter~to avoid short-range mutual coupling. The nanodisks are placed over a restricted distance of 5~$\mu$\meter~with respect to the center of the nanowire to mitigate the influence of the large contact electrodes on the local temperature perceived by the nanowire.

\subsection{\label{sec:level1}Electrical measurements}
We use the following procedure to measure the electrical properties of the Au nanowire. A function generator (Fran\c caise d'instrumentation) produces an AC voltage applied across the nanowire with an amplitude $V_{\rm AC}=20$~mV and a frequency $F=12.6$~k\hertz. This frequency is used as an external reference for a lock-in amplifier (UHFLI, Zurich Instrument). A current-to-voltage converter (DLPCA-200, Femto GmbH) detects the current flowing through the device, and the voltage output is sent to the lock-in amplifier as shown in Fig.~\ref{setup}.  The output of the lock-in provides a voltage signal proportional to the amplitude of the modulated current oscillating at  $F$. The differential resistance of the nanowire $R_{\rm nw}$ is then estimated by normalizing $V_{\rm AC}$ by the lock-in signal. We perform a 4-probe measurement~\cite{Okino2005,Gu2006,Walton2010} to remove the contributions of the macroscopic electrodes and their contacts in the estimation of the nanowire resistance $R_{\rm nw}$. The parasitic resistances are successively evaluated by applying the AC voltage $V_{\rm ij}$ to each pair of electrodes and we measure the corresponding lock-in output signals.  $i,j=1$ to 4 are the contact labels, as defined in Fig.~\ref{SEM}(a). The partial resistance of the sole nanowire writes  $R_{\rm L_{ij}}$ where $L_{ij}$ is the distance between the electrode labeled $i$ and the electrode labeled $j$.  The contact resistance of each electrodes is $R^c_{i}$. The total resistance is thus $R_ {ij}= R^c_{i}+R^c_{j} +R_{\rm L_{ij}}$. The measurement follows the sequence below:

\begin{align}
R_{\rm 13} = \frac{V_{\rm 13}}{I_{\rm 13}},       &&        
R_{\rm 23} = \frac{V_{\rm 23}}{I_{\rm 23}},
\end{align}
\begin{align*}
R_{\rm 14} = \frac{V_{\rm 14}}{I_{\rm 14}},       &&      
R_{\rm 24} = \frac{V_{\rm 24}}{I_{\rm 24}}.
\end{align*}

with this procedure, we estimate the resistance of the nanowire $R_{\rm nw}=R_{\rm L_{24}}$ corresponding to the central portion of the nanowire without the contact contributions using the relation~\cite{Gu2006}:

\begin{equation}
R_{\rm nw} = \frac{(R_{\rm 13} - R_{\rm 14}) + (R_{\rm 24} - R_{\rm 23})}{2}.
\label{Rnw}
\end{equation}
For the structures fabricated in this work, the resistance of the  nanowires may vary between processed samples. Otherwise noted, the nanowire considered in the following has a resistance $R_{\rm nw}=175$~\ohm, which corresponds to a metal resistivity  for the evaporated layers of $\rho = R_{\rm nw}\times L_{\rm nw}/S=6.5$ $\mu$\ohm~c\metre~where $S$ is the section of the nanowire.  This value is consistent with the gold's resistivity reported from mesoscopic structures~\cite{VanAttekum1984}.

During the optical measurements discussed below, the bias $V_{\rm AC}$ is applied across the contacts labeled 1 and 3 in a 2-probe configuration. The differential resistance $R$ estimated from the output of the lock-in-amplifier is then equal to the sum of the resistance of the two unknown electrical contacts $R^{\rm c}_{1}$ and $R^{\rm c}_{3}$ and the resistances $R_{\rm L_{14}}$, $R_{\rm nw}$ and $R_{\rm L_{23}}$ corresponding to the relevant portions of the nanowire. Assuming a constant resistivity $\rho$,

\begin{equation}
\begin{split}
R &= R_{\rm 13}\\
&= R^{\rm c}_{1} + R^{\rm c}_{3} + R_{\rm L_{14}} + R_{\rm nw} + R_{\rm L_{23}}\\
&= R^{\rm c}_{1} + R^{\rm c}_{3} + \frac{\rho}{S}(L_{\rm 14} + L_{\rm 23})+R_{\rm nw}.
\end{split}
\label{dR}
\end{equation}
$L_{\rm 14}$ and $L_{\rm 23}$ are the length between the contacts 1--4 and  2--3, respectively. Knowing the resistance of the nanowire $R_{\rm nw}=175$~\ohm~(Eq.~\ref{Rnw}), the normalized output of the lock-in $R_{\rm 13} $ and $\rho(L_{\rm 14} + L_{\rm 23})/S=63$~\ohm, the resistance of the contacts can be inferred $R^{\rm c}_{1}$ +  $R^{\rm c}_{3}=409$~\ohm. The total contribution of the connection to the resistance amounts to 472~\ohm.  This value is subtracted from the resistance estimated from the 2-probe lock-in measurement.

\subsection{\label{sec:level1} Local optical heating of the gold nanodisks}
\begin{figure}[h]
\includegraphics[width=10cm]{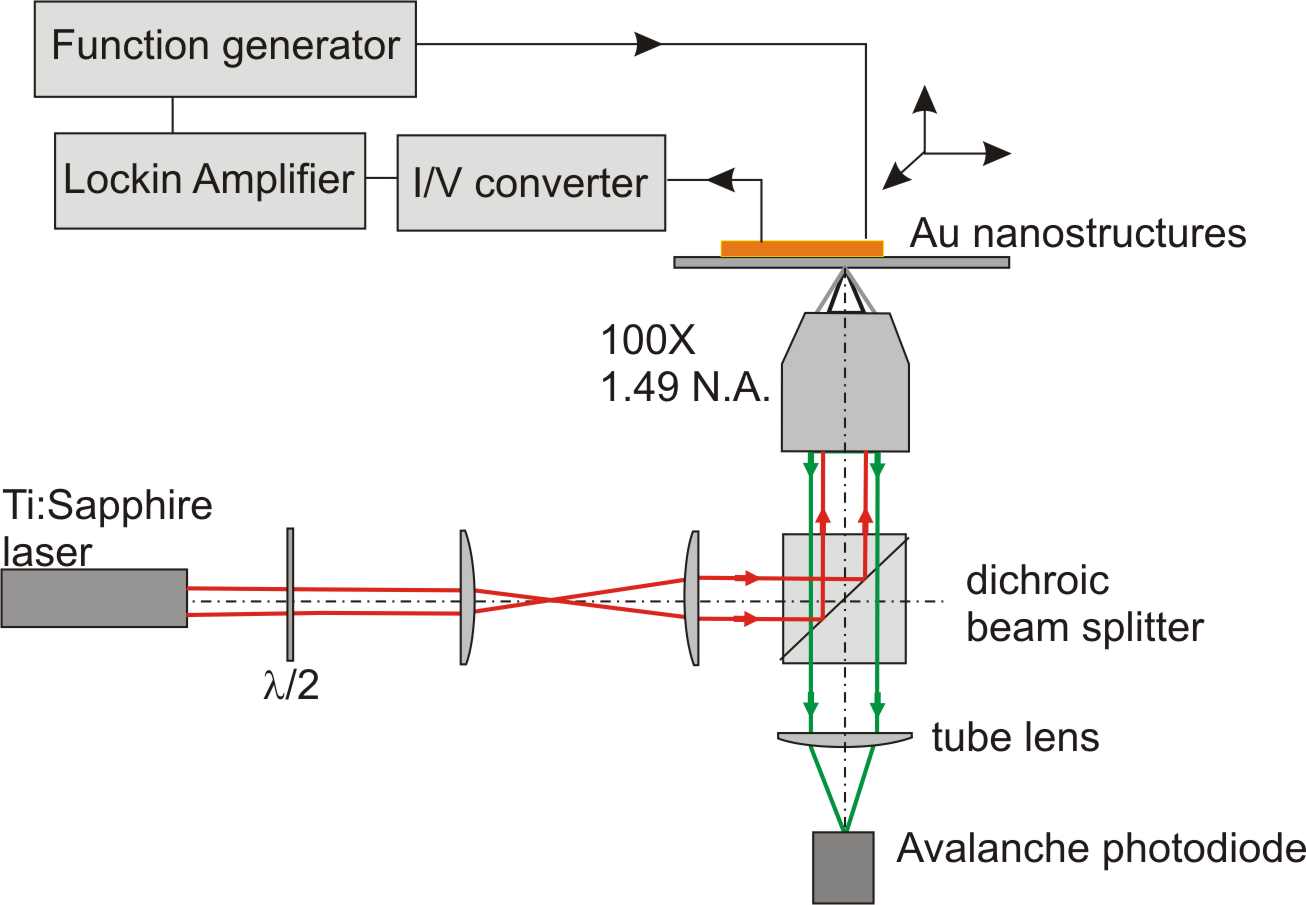} 
\caption{Sketch of the experimental setup. The sample is scanned in the focus of a pulsed laser emitting 180~f\second~pulses with a repetition rate of 80~M\hertz~ centered at a wavelength of 810~n\meter. The nonlinear optical response (two-photon luminescence) is detected by an avalanche photodiode. The influence of the optical excitation on the electronic transport of the nanowire is recorded by a current-to-voltage converter feeding a lock-in amplifier. The lock-in amplifier is synced to the frequency of the AC bias applied to the nanowire by the function generator.}
\label{setup}
\end{figure}

Figure~\ref{setup} shows a schematic of the setup used in this study. The nanodisks are excited by a tightly focused Ti:Sapphire laser beam emitting a 180~f\second~pulses at a repetition rate of 80~M\hertz~(Chameleon, Coherent). The wavelength is kept constant at 810~n\meter. The sample is placed on an inverted optical microscope (Eclipse, Nikon) equipped with a two-axis piezoelectric actuator (NanoLP-100, MadCity Labs) raster scanning the sample in the focal region. A 100$\times$ objective with a numerical aperture of 1.49 focuses the laser beam in a diffraction limited spot and collects the nonlinear two-photon luminescence produced by the Au nanostructures~\cite{Bouhelier2003,bouhelier05PRL}. The nonlinear signal is detected by an avalanche photodiode (Excelitas). The luminescence is used for imaging and analysis purposes. The average laser intensity at the sample is 420~k\watt~c\meter$^{-2}$ and the polarization of the focused laser beam is linearly polarized perpendicularly to the nanowire in order to homogenize the response along the object.  We verified that the conclusions described below do not depend on the use of a pulsed laser. Leaving aside the nonlinear activity, similar results were consistently obtained for a constant-wave laser of the same power and emitting at 785~nm.

\begin{figure}[h]
\includegraphics[width=15cm]{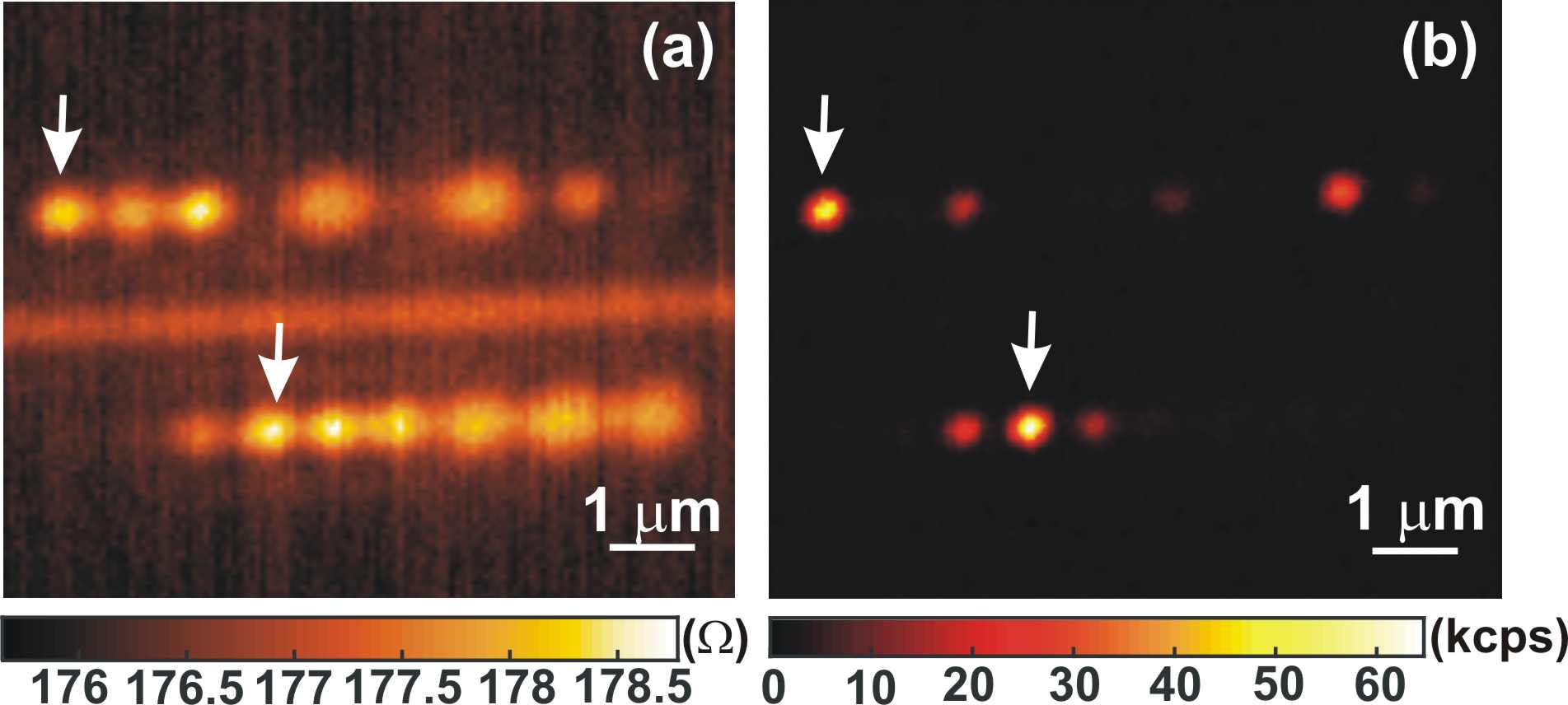} 
\caption{(a) Image of the variation of the nanowire's resistance $\Delta R_{\rm nw}$ for a lateral scan of the sample in the focal spot of the laser. Heat generated by the laser on the different part of the sample provides an increase in of the nanowire's resistance. (b) Simultaneously acquired nonlinear two-photon luminescence emitted by the Au structures. The nanodisks in resonance with the laser (arrows) are inducing the largest variation of the resistance and the strongest nonlinear photoluminescence.  }
\label{confocalRTPL}
\end{figure}

Upon local illumination, the absorbed energy is converted to heat. For the irradiated nanodisks, the heat transported through the substrate may reach the nanowire. The amount of heat perceived by the distant nanowire induces a drop of the electronic transport, which depends linearly on the nanowire temperature~\cite{piróth2008fundamentals}. The differential resistance and the nonlinear signal emitted by the structures are recorded for each sample position in the focus of the laser to form the images depicted in Fig.~\ref{confocalRTPL}.  In Fig.~\ref{confocalRTPL}(a) a positive variation of the resistance of the nanowire is clearly observed when the metal structures are optically excited. The contrast is constant when the nanowire is directly excited by the laser with a modest increase of the resistance. However, when the nanodisks are scanned through the focus, the variation of the nanowire's resistance $\Delta R_{\rm nw}$ becomes dependent on the diameter of the nanodisks.  Variations of $\Delta R_{\rm nw}$ up to 1.4~\% are recorded for the $\sim$210~n\meter~nanodisks as indicated by the arrows. These nanodisks are the elements circled in Fig.~\ref{SEM}.  For nearly all diameters, the variation of the nanowire's resistance is larger when the laser irradiates the isolated nanodisks than for a direct illumination of the nanowire itself. 

\begin{figure}[h]
\includegraphics[width=15cm]{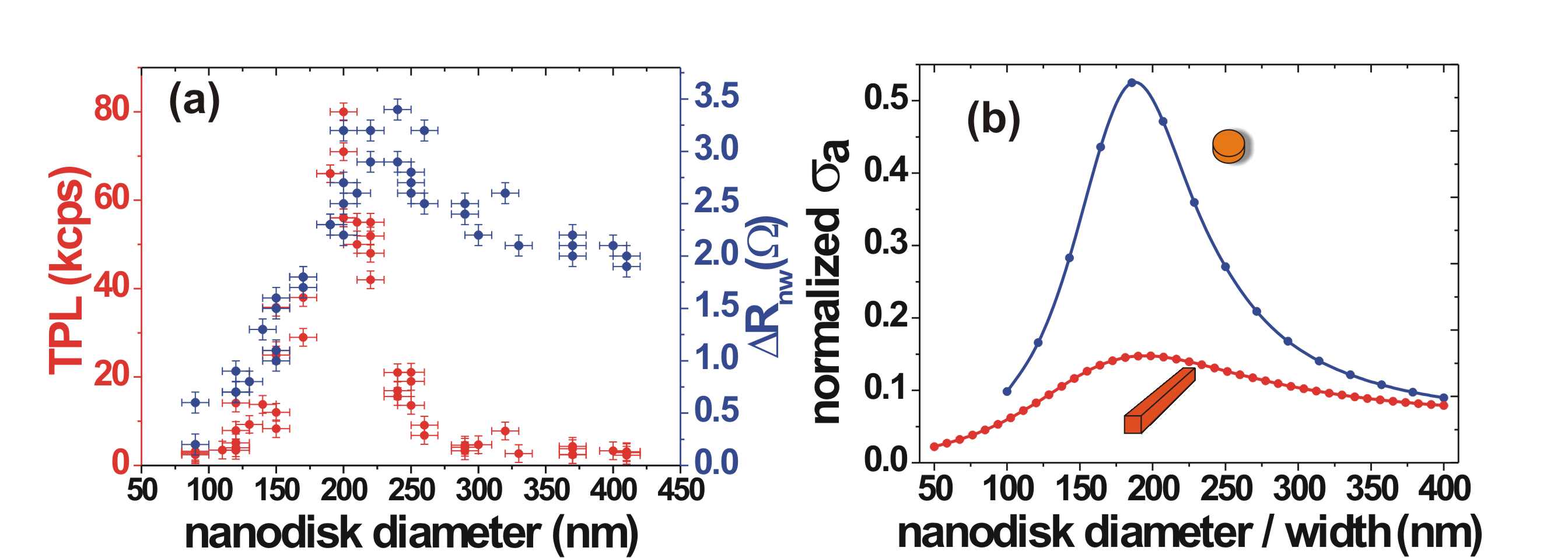} 
\caption{(a) Evolution of differential resistance and the two-photon luminescence (TPL) with nanodisk diameter. A plasmon resonance is clearly identified for nanodisks with a diameter of 210~nm. The graph gathers measurements obtained from three different samples. (b) Blue curve: Simulated evolution of the absorption efficiency of a Au nanodisk placed on a glass substrate as a function of its diameter. The simulated resonant absorption occurs for $\sim$200~nm nanodisk. Red curve: Absoprtion efficiency for a nanowire as a function of its width. The polarization is perpendicular to the main axis of the nanowire. }
\label{diameter}
\end{figure}

Heat transferred during the optical excitation of nanodisks depends directly on their absorbing capacity, which is enhanced at the surface plasmon resonance. The occurrence of the resonance is readily identified in the nonlinear two-photon luminescence produced by the nanodisks~ \cite{bouhelier05PRL,quidant08}.  as illustrated in the confocal map of Fig.~\ref{confocalRTPL}(b). The strongest response is observed for the  210~n\meter~nanodisk (arrows). The red curve in Fig.~\ref{diameter} (a) shows the nonlinear photoluminescence dependence with the size of the nanodisks obtained from three different samples. At the laser wavelength used in this study, the $\sim$ 210~n\meter~nanodisk sustains a marked resonance. The largest $\Delta R_{\rm nw}$ in Figure~\ref{confocalRTPL}(a) is also observed for this diameter.

The resonance is numerically confirmed by the evolution of the simulated scattering absorption cross-section $\sigma_{\rm a}$ of a nanodisk as a function of its diameter. The absorption cross-section of the Au structures writes:
\begin{equation}
\sigma_a(\omega)=\frac{P_a(\omega)}{\psi_i(\omega)}.
\end{equation}

$\sigma_a(\omega)$ is defined for an incident monochromatic plane wave of frequency $\omega$ as  the ratio of the absorbed power

\begin{equation}
P_a(\omega)=\frac{\omega}{2} \int_v \left| E \right|^2 Im(\varepsilon) dV
\end{equation}
by the incident flux
\begin{equation}
\psi_i(\omega) =\frac{1}{2} \left(\frac{\epsilon_0}{\mu_0}\right)^{1/2}\left| E \right|^2.
\end{equation}

Here $\varepsilon$ denotes the electric permittivity of the absorber of volume $V$, $E$ is the local electric field and $\epsilon_0$ and $\mu_0$ are the vacuum permittivity and permeability, respectively.  The absorption cross-sections of the nanodisks and of the nanowire are calculated by solving Maxwell's equations with a finite-element method combined to an analytical description of the field scattered at the interface of the substrate~\cite{FEM}. The absorption cross-section is represented by the blue curve in Fig.~\ref{diameter} (b) after a normalization by the surface of the absorbers. The presence of the glass substrate is taken into account in the calculations. The simulation clearly indicates an enhanced absorption efficiency for a particle diameter $\sim$200~n\meter. The red curve in Fig.~\ref{diameter} (b) is the calculated absorption efficiency for a nanowire with varying widths. The structure is assumed invariant along the nanowire axis and the calculation is performed for a linear polarization aligned with the short axis. For the experimental width considered here (75~n\meter),  $\sigma_{\rm a}$ is $\sim 5$ times weaker than for a resonant nanodisk (blue curve) and explains the dimmer contrast observed for a direct illumination of the nanowire in Fig.~\ref{confocalRTPL}(a) despite the fact that the nanodisks are laterally shifted by 1~$\mu$m. These consistent results confirm the unambiguous role of the surface plasmon resonance at generating an absorption enhancement and the subsequent rise of the temperature-dependent resistance of the nanowire~\cite{Herzog2014}.

Interestingly, the variation of the resistance follows a different trend than the nonlinear photoluminescence as illustrated by the blue curve in Fig.~\ref{diameter}(a). $\Delta R_{\rm nw}$ increases to reach a maximum at 3.5~\ohm~ for $\sim$220~n\meter~nanodisks and then reduces to reach a plateau at 2~\ohm~for diameter larger than 350~n\meter.  This diameter corresponds to the approximate size of the focal spot.  The evolution of $\Delta R_{\rm nw}$ is thus the convolution of a profile related to the plasmon resonance with a size dependence related to the saturation of the optical absorption for nanodisks larger than the focal spot.

\begin{figure}[h]
\includegraphics[width=15cm]{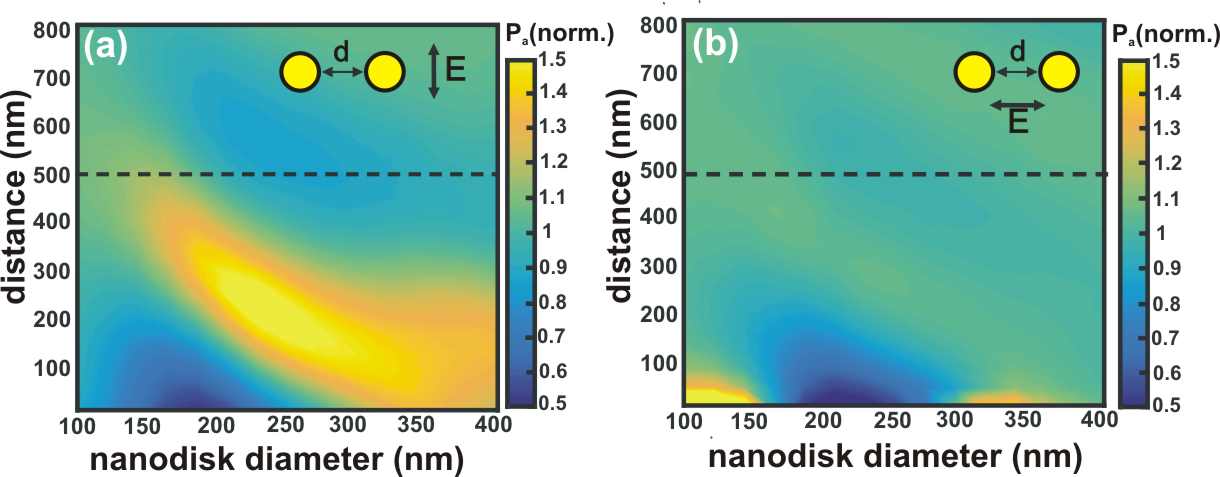} 
\caption{(a) and (b) Dependence of the power absorbed by a nanodisk in the presence of a second one for varying edge-to-edge distances $d$ and disk diameters. The maps are calculated for a TE polarization and a TM polarization, respectively. The absorbed power is normalized to the absorption of a single isolated nanodisk. The excitation is a plane wave normally incident to the interface with a wavelength $\lambda=810$~nm. The dashed line corresponds the minimum experimental distance between the nanodisks in Fig.~\ref{SEM}. }
\label{distancevsdiameter}
\end{figure}

We verified numerically that mutual interactions between nanodisks can be neglected with the interparticle distance considered in this work. 
Figure~\ref{distancevsdiameter} describes the evolution of the optical absorption $\sigma_a$ of an excited nanodisk in the presence of a second gold nanodisk placed at a distance $d$ from the first, and that for two crossed polarizations. The dashed line is the edge-to-edge distance used in the experiment. Interestingly, the calculations predict that the absorption of a nanodisk can be enhanced or reduced by adjusting either of the three parameters. However, for the case considered here the optical absorption and therefore all related phenomena, such as the temperature dissipated in the substrate, are not disturbed by the presence of other nanodisks.

\subsection{\label{sec:level1}Temperature calibration}
\begin{figure}[h]
\includegraphics[width=15cm]{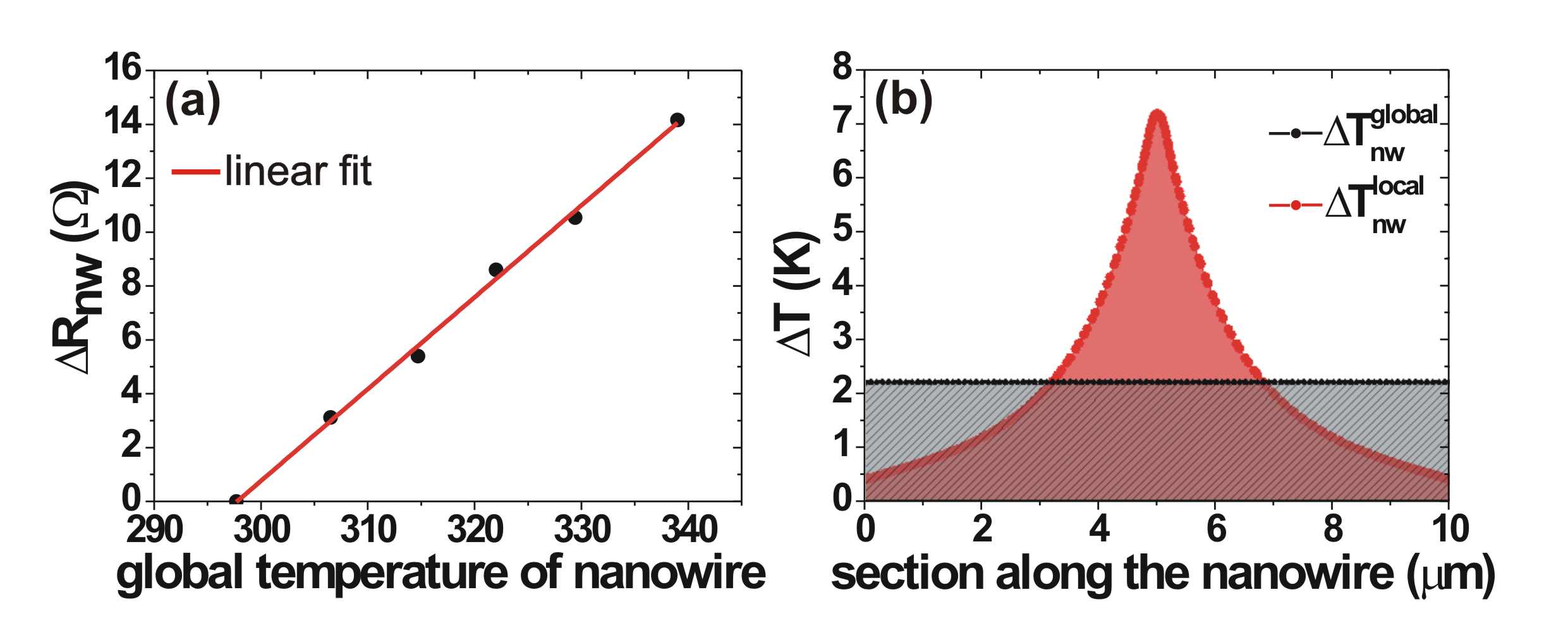} 
\caption{(a) Evolution of nanowire resistance $R_{\rm nw}$ with a global increase of the sample temperature. The entire sample is heated with a Peltier Module. (b) Black curve: longitudinal distribution of the temperature along the nanowire estimated from the $dR_{\rm nw}/dT_{\rm global}$ coefficient deduced from (a) and the $\Delta R_{\rm nw}$ recorded by the lock-in amplifier during laser illumination. Red curve: simulated distribution of the temperature along the nanowire for a local heating element centered on the nanowire. The maximum temperature is computed to set the surface integral of the red curve equals to that of the black curve.}
\label{calibration}
\end{figure}

On order to estimate the rise of the nanowire's temperature, we calibrate the temperature-dependent resistance of the nanowire  $R_{\rm nw} (T)$ by placing the sample on a controlled heating module in the line with the reports by Herzog and co-workers~\citep{Herzog2014} and Stolz \textit{et al}~\cite{Stolz2014}. Figure~\ref{calibration}(a) shows the linear variation of the resistance as a function of the temperature $T_{\rm global}$ heating the entire sample. The resulting coefficient $dR_{\rm nw}/dT_{\rm global}=0.41$~\ohm~\kelvin$^{-1}$ depends on the resistivity of the nanowire and its characteristic dimensions. The variation $\Delta R_{\rm nw}$ recorded during the lateral scan of the nanowire in the focal spot (Fig.~\ref{confocalRTPL}(a)) may be linked to a first approximation to the temperature variation $\Delta T_{\rm global}=\Delta T^{\rm global}_{\rm nw}$ evenly distributed over the nanowire. The black curve in Fig.~\ref{calibration}(b) is the inferred temperature elevation of the nanowire $\Delta T^{\rm global}_{\rm nw}=2.3$~\kelvin~considering the rise of resistance $\Delta R_{\rm nw}$ measured in Fig.~\ref{confocalRTPL}(a) when the laser irradiates directly the nanowire.  

This approximation is however not representative of a local excitation induced by a tightly focused laser beam. To motivate this point, we show in Fig.~\ref{calibration}(b) a finite-element simulation (Comsol multiphysics) of the longitudinal distribution of $\Delta T^{\rm local}_{\rm nw}$ produced by a simulated heat source located at the center of the nanowire and with a dimension corresponding to the focal spot. The geometry of the nanowire and of the nanodisks used in the computation is that of the experimental structures without the adhesion layer. Two 500~n\meter$\times~20~\mu\meter$~gold pads mimicking the electrodes are placed at the extremities of the nanowire. The nanowire, the nanodisks and the pads are deposited on a $50~\mu\meter\times50~\mu\meter\times 50~\mu\meter$ cube of glass. The material properties used in the calculations are for the thermal capacity; $C^{\rm Au}=129~\joule~\kilo\gram^{-1}~\kelvin^{-1}$ and $C^{\rm glass}=730~\joule~\kilo\gram^{-1}~\kelvin^{-1}$, for the thermal conductivity: $\kappa^{\rm Au} = 317~\watt~\meter^{-1}\kelvin^{-1}$ and $\kappa^{\rm glass}=1.4~\watt~\meter^{-1}~\kelvin^{-1}$, and for the electrical conductivity: $\sigma^{\rm Au}=45.6\times10^6~\siemens~\meter^{-1}$ and   $\sigma^{\rm glass}=1\times10^{-14}~\siemens~\meter^{-1}$. The initial temperature at the boundaries of the calculation window is 296.15~\kelvin. The surface emissivity for Au is $\epsilon^{\rm Au}=0.02$, and for glass $\epsilon^{\rm glass}=0.94$. 

To estimate the local temperature of the heat source we equalize the resistance variation due to an homogeneous heating with a rise of temperature $\Delta T^{\rm global}_{\rm nw}$ using the Peltier module and the resistance change by a local heating with a temperature $\Delta T^{\rm local}_{\rm nw}(x)$. $x$ is the coordinate along the nanowire length $L_{\rm nw}$. For a temperature-dependent resistivity $\rho(T)$ this equality writes

\begin{equation}
\Delta R_{\rm nw} = \frac{\rho(\Delta T^{\rm global}_{\rm nw}) L_{\rm nw}}{S}=\int_0^{L_{\rm nw}}\frac{\rho(\Delta T^{\rm local}_{\rm nw}(x))}{S} dx.
\label{equal_resistance}
\end{equation}
From the linear relationship between $R_{\rm nw}$ and $T^{\rm global}_{\rm nw}$ depicted in Fig.~\ref{calibration}(a), Eq.~\ref{equal_resistance} is satisfied if the  surface integrals of the profiles $\Delta T^{\rm local}_{\rm nw}$ and $\Delta T^{\rm global}_{\rm nw}$ along the nanowire are equal. We thus compute the maximum local temperature $\Delta T^{\rm local,max}_{\rm nw}$ to satisfy the equality. In Fig.~\ref{calibration} (b),  we set $\Delta T^{\rm local,max}_{\rm nw}=7.2$~\kelvin~to have identical areas under the two curves. For the same change of resistance the maximum temperature reached by the nanowire $\Delta T^{\rm local,max}_{\rm nw}$ is thus three times higher for a diffraction-limited local heat source than for a global heating of the system. This approach is valid for a local heating of the nanowire, but would fail at describing a situation where a local rise of the temperature would be compensated by a local cooling of the system. In this particular case, $R_{\rm nw}$ would remain unchanged.

\begin{figure}[h]
\includegraphics[width=15cm]{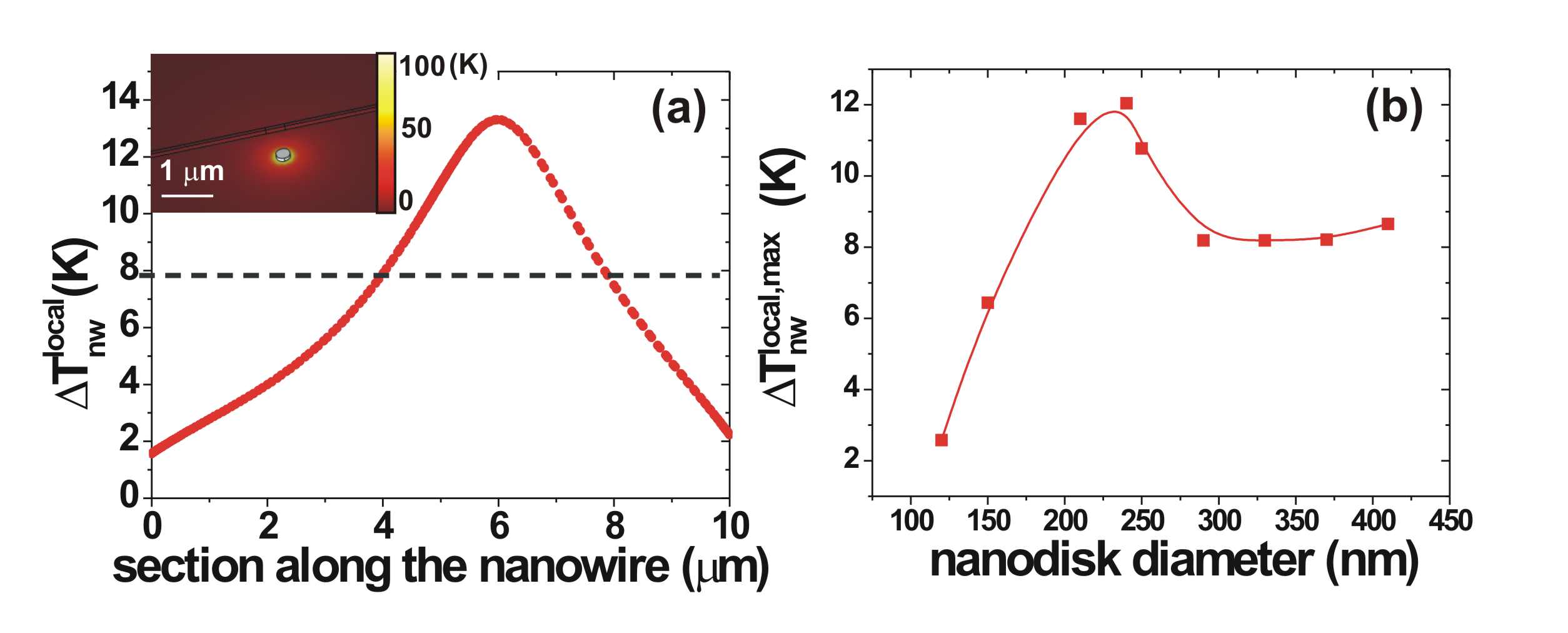} 
\caption{(a) Estimated temperature gradient along the nanowire when a 210~nm diameter nanodisk is heated at a distance of 1~$\mu$m and introduces a 2.8~\ohm~rise of the nanowire's resistance. Dashed line: global temperature required to change the resistance by the same amount.  Inset: Spatial distribution of the simulated temperature in the computed system. (b) Maximum temperature deduced from the experimental variation of the nanowire's resistance as a function of the nanodisk's diameter.}
\label{Tdisk}
\end{figure}

This approach is applied to estimate the temperature distribution along the nanowire for a laser excitation of the gold nanodisks. Under this configuration heat must diffuse through the substrate to impact the nanowire's resistance. An example of the procedure is shown in Fig.~\ref{Tdisk}(a) for the resonant disk identified in the $\Delta R_{\rm nw}$ map of Fig.~\ref{confocalRTPL}(a). The curve is simulated using the following steps. We first determine the experimental value of $\Delta R_{\rm nw}$ of the nanowire when the nanodisk is irradiated. For the example considered here $\Delta R_{\rm nw}=2.8$~\ohm. Using the global temperature coefficient $dR_{\rm nw}/dT_{\rm global}=0.41$~\ohm~\kelvin$^{-1}$, we estimate what would be the necessary global temperature required to increase the resistance by 2.8~\ohm, that is $\Delta T_{\rm global}=6.9$~\kelvin. This hypothetical temperature is homogeneously distributed along the nanowire (dashed curve in Fig.~\ref{Tdisk}(a)). To take into account the local illumination of the nanodisk by the focused laser beam, we compute the two-dimensional distribution of the temperature in the system considering a heated nanodisk (see inset in Fig.~\ref{Tdisk}(a)). A profile of the temperature distributed along the nanowire is illustrated by the red curve in Fig.~\ref{Tdisk}(a).  We then adjust the temperature of the heat source in the finite-element simulation of the system to obtain equal surface integrals between the temperature profiles perceived by the nanowire. For a temperature rise of the nanodisk of 108~\kelvin, the surfaces behind the two curves in Fig.~\ref{Tdisk}(a) are identical and the maximum temperature rise at the nanowire is about $\Delta T^{\rm local,max}_{\rm nw}=13$~\kelvin. Conducting this analysis for the range of diameters considered in this study, we show in Fig.~\ref{Tdisk}(b) that $\Delta T^{\rm local, max}_{\rm nw}$ remains significant even for optically pumped nanodisks laterally shifted by 1~$\mu$m.

In the following, we estimate the rise of the temperature in the nanowire purely from a simulation stand point. The objective is to verify the reliability of the temperature variations deduced from the approach exposed above.  To do so, the problem is solved by a finite-element analysis of the combined physical processes at play.  The temperature of the nanodisk is no longer derived from the variation of the nanowire's resistance. It is directly computed from the electromagnetic problem consisting at calculating the power absorbed by the Au nanodisks upon illumination of a simulated focused beam. 

\begin{figure}[h]
\includegraphics[width=15cm]{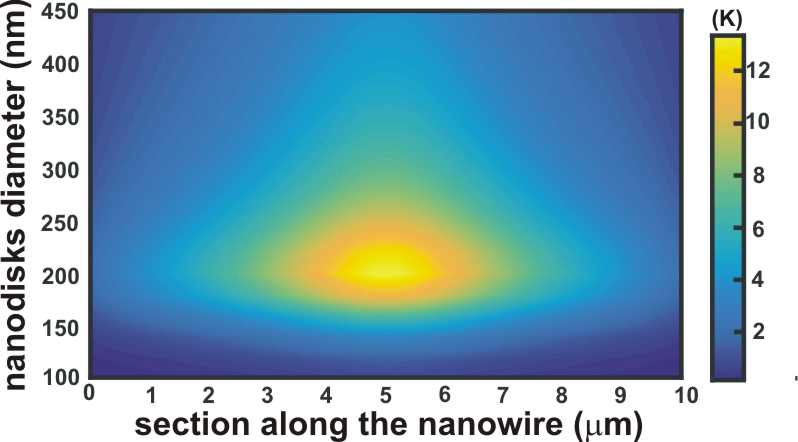} 
\caption{Computed temperature variation along the nanowire for varying nanodisk diameters placed at a distance of 1~$\mu$\meter. The nanodisks are the source of heat. The graph is calculated by a finite-element analysis taking into account the experimental conditions (focused beam, incident power and size of the structures). }
\label{computedTnw}
\end{figure}

We compute the steady-state longitudinal distribution of the temperature along the nanowire for the different nanodisks considered experimentally. The results are reported in Fig.~\ref{computedTnw}.  The highest computed temperature variation $\Delta T^{\rm local, max}_{\rm nw}=12$~\kelvin~ is in very good agreement with the temperature determined from the fluctuations of nanowire electron transport (Fig.~\ref{Tdisk}(b)). Furthermore, the full calculation reproduces the plateau observed for large diameters in Fig.~\ref{Tdisk}(b). These similarities demonstrate the consistency of the resistance analysis used to deduce the local elevation of the temperature $\Delta T^{\rm local, max}_{\rm nw}$.

For diameters exceeding the size of the laser beam, the simulated $\Delta T^{\rm local,max}_{\rm nw}$ is  a factor two lower than the experimentally inferred value. The profile of the temperature along the nanowire used for the calculation of $\Delta T^{\rm local,max}_{\rm nw}$ is obtained by the diffusion of a homogeneous heat source through the glass substrate. This approximation is correct when the plasmonic nanodisk is smaller than the laser beam. For larger diameters, the optical excitation of nanodisk can no longer be approximated to a homogeneous heat source and is the probable source of the discrepancy.

\section{\label{sec:level1}Long range heat transfer}

Thus far we considered illuminated nanodisks laterally displaced by 1~$\mu$m from the nanowire and estimated a 13~\kelvin~rise of the local temperature when resonant nanodisks are illuminated. In the following we investigate the distance dependence of the heat transfer via the glass substrate. 

\begin{figure}
\includegraphics[width=15cm]{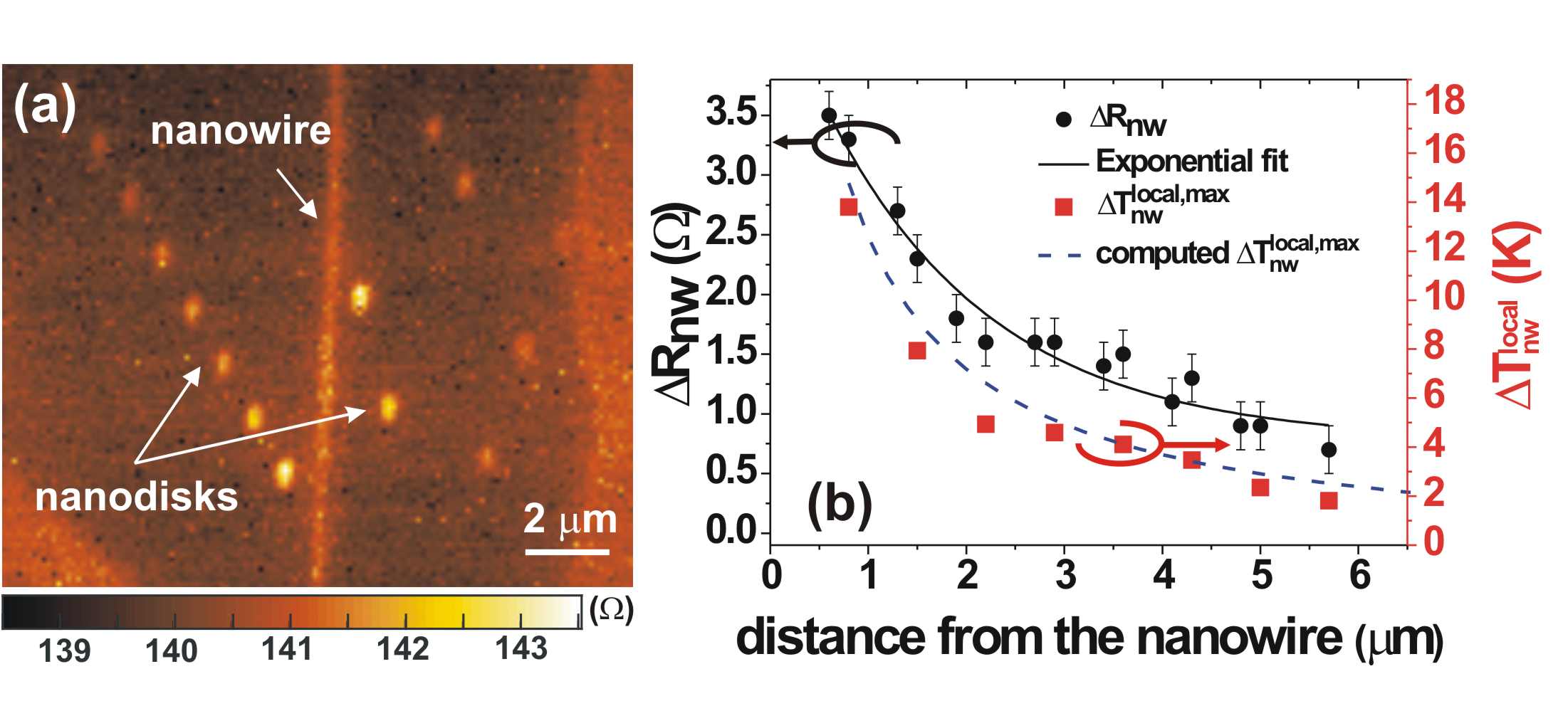} 
\caption{ (a) Two-dimensional evolution of the nanowire's resistance when the structure is scanned through the focus. The resistance increases whenever the laser is heating the nanostructures in the scanned area. On the left side of the nanowire, the distance of the resonant nanodisks is increasing step wise. On the right side, nanodisks are randomly shifted from the nanowire. (b) Left axis: (circle) variation of the nanowire's resistance $\Delta R_{\rm nw}$ as a function of the distance between the nanodisks and the nanowire. Solid black line: Best-fit exponential curve to the experimental data points with a characteristic decay length of 1.7~$\mu$m. Right axis: (square) Estimated variation of the maximum temperature perceived by the nanowire. Blue dashed line: computed maximum variation of the temperature. }
\label{distance}
\end{figure}

In this experiment, we place resonant nanodisks at distances from the nanowire ranging from  800~nm to 6~$\mu$m. The study is conducted on a nanowire with a resistance $R_{\rm nw}=138$~\ohm~ and we maintain constant the optical conditions. Figure~\ref{distance}(a) is a two-dimensional distribution of the resistance $R_{\rm nw}$ when the sample is laterally scanned through the focus. We observe the same contrast as in Fig.~\ref{confocalRTPL}(a) with an increase of the nanowire's resistance when the Au structures are heated by the laser. We plot in the left axis of Fig.~\ref{distance}(b) the variation of the resistance $\Delta R_{\rm nw}$ as a function of the lateral displacement of the nanodisks. The heat generated by the laser incident on a nanodisk remotely placed as far as 6~$\mu$m and diffusing through the glass substrate imparts a measurable rise of the nanowire resistance. The exponential fit to the data provides a decaying diffusion with a characteristic length of 1.7~$\mu$m. 

Following the procedure described above, we calculate  the maximum temperature variation $\Delta T^{\rm local,max}_{\rm nw}$ via the calibration coefficient $dR_{\rm nw}/dT_{\rm global}$ determined by a global heating (Fig.~\ref{calibration}(a)) and the spatial distribution of temperature determined by the finite-element analysis (Fig.~\ref{calibration}(a)). The maximum temperature $\Delta T^{\rm local,max}_{\rm nw}$ is plotted in the right axis of Fig.~\ref{distance}(b). At a distance of 800~nm from the nanowire, the excitation of a resonant structure rises the nanowire's temperature by about  14~\kelvin. For the same nanodisk placed at 6~$\mu$m, $\Delta T^{\rm local,max}_{\rm nw}$ drops to 2~\kelvin. The dashed blue curve is the calculated $\Delta T^{\rm local,max}_{\rm nw}$ and is in very good agreement with the points (square) deduced from the experimental $\Delta R_{\rm nw}$.

\section{\label{sec:level1}Conclusions}

We have investigated the effect of a remote resonant plasmonic photothermal excitation on the electronic transport of a Au nanowire. Upon focusing light on isolated Au nanodisks, the heat locally produced diffuses through the substrate and induces a change of the nanowire's resistance. We estimate the temperature rise at the nanowire by taking into account the local nature of the excitation, the experimental data and finite-element calculations. We estimate a local temperature rise at the nanowire of 13~\kelvin~for a laser intensity of 420~k\watt~c\meter$^{-2}$ focused on a resonant plasmonic nanoparticle located 1~$\mu$m away from the nanowire. The photothermal excitation of an electrically contacted metal nanowire resting on the glass surface is usually considered negligible. While the macroscopic electrodes connected to the nanowire act as heat sinks and the physical contact to the substrate contributes to an efficient cooling of the nanowire by thermal conduction~\cite{kopp12}, we determine a maximum temperature significantly exceeding the average temperature inferred from a global calibration of the temperature-dependent resistivity of the nanowire~\cite{Ward2010,Herzog2014,Stolz2014}. Interestingly, we demonstrate that the rise of temperature for a direct optical excitation of the nanowire is weaker than for a laterally displaced isolated resonant nanoparticle. The thermal diffusion through the glass substrate follows an exponential decay of about 2~$\mu$m, and illuminated nanodisks located as far as 6~$\mu$m generate a temperature variation of 2~\kelvin~at the nanowire.  While all the experiments are conducted in the steady-state regime, the protocol can be readily extended to a time-dependent thermoplasmonic modulation of the electronic transport of the nanowire.  To conclude, we demonstrate that relatively simple electrical measurement can be implemented to remotely probe the plasmon-induced photothermal activity of an illuminated device. By extension, measuring the resistance variations of a network of such heat probes may enable a diffusion-based thermal tomography of a local heat source. We believe that thermal diffusion through the substrate may be mitigated or conversely enhanced by a careful control of the electromagnetic interaction between an ensemble of nanoparticles. In line with this concluding remark, a phononic engineering~\cite{Maldovan13} of the substrate's local thermal conductivity would also offer a leverage to control the directional flow of the heat produced by the resonant absorption.

\section{Acknowledgments}
The  research  leading  to  these  results  has  received funding  from  the  European  Research  Council  under  the European Community's Seventh Framework Program FP7/ 2007-2013 Grant Agreement No. 306772. This project is in cooperation with the Labex ACTION (Contract No. ANR-11-LABX-01-01) and the regional PARI programs. The authors are grateful to O. Demichel for valuable discussions.

%

\end{document}